%Paper: hep-th/9302009
%From: CRESCIMANNO@PIERRE.MIT.EDU
%Date: Tue, 2 Feb 1993 22:06:15 -0500 (EST)

\def\ll{\left\langle}
\def\rr{\right\rangle}

\magnification=1200
\hoffset=-.1in
\voffset=-.2in

\vsize=7.5in
\hsize=5.6in
\tolerance 10000

\baselineskip 12pt plus 1pt minus 1pt
\topinsert
%\hfill {\bfmone PRELIMINARY}
\bigskip
\endinsert
\pageno=0
\centerline{\bf EQUILIBRIUM TWO-DIMENSIONAL}
\smallskip
\centerline{{\bf DILATONIC SPACETIMES}\footnote{*}{This
work is supported in part by funds
provided by the U. S. Department of Energy (D.O.E.) under contract
\#DE-AC02-76ER03069, and by the Division of Applied Mathematics of the U.S.
Department of Energy under contract \#DE-FG02-88ER24066}}
\vskip 24pt
\centerline{Michael Crescimanno}
\vskip 12pt
\centerline{\it Center for Theoretical Physics}
\centerline{\it Laboratory for Nuclear Science}
\centerline{\it and Department of Physics}
\centerline{\it Massachusetts Institute of Technology}
\centerline{\it Cambridge, Massachusetts\ \ 02139\ \ \ U.S.A.}
\vskip 1.5in
\centerline{Submitted to: {\it Physical Review D\/}}
\vfill
\centerline{ Typeset in $\TeX$ by Roger L. Gilson}
\vskip -12pt
\noindent CTP\#2174\hfill January 1993
\eject
\baselineskip 24pt plus 2pt minus 2pt
\centerline{\bf ABSTRACT}
\medskip

We study two-dimensional dilaton gravity coupled to
massless scalar fields and look for time-independent
solutions.  In addition to the well-known black hole,
we find another class of solutions that may be understood
as the black hole in equilibrium with a radiation bath.
We claim that there is a solution that is qualitatively
unchanged after including Hawking radiation and back-reaction
and is geodesically complete. We compute the thermodynamics
of these spacetimes and their mass. We end with a
brief discussion of the linear response about these solutions,
its significance to stability and noise, and a speculation
regarding the endpoint of Hawking evaporation in four dimensions.

\vfill
\eject
\noindent{\bf I.\quad INTRODUCTION}
\medskip
\nobreak
Two-dimensional gravity has emerged as a useful toy model for addressing some
basic philosophic questions about semiclassical and quantum gravity.
Dilation gravity is the low-energy effective theory of string theory and so
solutions of this theory arise naturally as the $\sigma$-models backgrounds
associated to conformal field theories.  An example of these are the coset
models$^{1,\,2,\,3}$ among which was found two-dimensional $\sigma$-models,
including that of the black hole,$^{4,\,5,\,6}$
which is a solution of two-dimensional
dilaton gravity with a Minkowski signature.
Besides just the graviton and dilaton,  a two-dimensional gravity also
arises naturally from dimensional reductions of ordinary Einstein gravity in
four and five dimensions (see Refs.~[7,8,9]).  Realistic string theory should
also have low-energy ``matter'' fields. Callan, {\it et al.\/}$^{10}$ showed
that including massless scalar fields to dilaton gravity in two dimensions
yields a toy model rich with phenomena such as gravitational collapse and
Hawking radiations.  This model has subsequently been explored by many
authors, who addressed a host of questions ranging from the structure of the
singularity,$^{11}$ of the Hawking radiation and its back
reaction on the metric,$^{12,\,13}$ information loss in black hole
evaporation,$^{14,\,15,\,16}$ and approaches to the full quantum
theory.$^{17-22}$  There are several very readable reviews on this subject;
see for example Refs.~[8,23].

Although two-dimensional dilaton gravity is somewhat different than ordinary
Einstein--Hilbert gravity, there are many features of the model
that are generic to other dimensions and other metric theories of gravitation.
 Certainly one difficulty in trying to answer some of the above questions is
that they concern (inherently) non-equilibrium phenomena.  Since the Green's
functions, that arise naturally when studying quantum fields in these
backgrounds are
typically equilibrium Green's functions they are often not suitable for
addressing these non-equilibrium questions beyond low orders in perturbation
theory.

There are, however, many interesting questions to be answered about the
equilibrium properties of semiclassical gravity, such as the nature of the
equilibrium state and the power spectrum of noise in the spacetime itself.
These issues have been somewhat neglected, as they do not allow one to
directly answer the interesting non-equilibrium questions posed above.
Alternatively, these notions are computable and rigorously defined.

Here we study (static) solutions to two-dimensional dilatonic
gravity coupled to matter and try to answer some of the equilibrium questions
above.  We will find that although the ordinary dilaton black hole is not an
equilibrium solution (that is, semiclasically with realistic boundary
conditions or once back-reaction is included) there does exist a
solution which is non-singular, geodesically complete and static.  This
solution seems to persist even with the back-reaction included.  It is
interpreted as the black hole in equilibrium with a heat bath of radiation.
We then compute its thermodynamic properties and describe the
fluctuations about this solution.
The idea of studying the two-dimensional dilatonic black hole in
equilibrium with a radiation
bath is not new (see for example the discussion in Refs.~[24--26,16]). In some
earlier works these solutions were discarded on the basis that they are not
finite energy solutions.  This is trivially true for black holes in a
radiation bath  in any open spacetime and so here we accept this as a simple
fact and instead focus on what we feel are more pressing physics issues. At
the end we remark on whether something like these solutions may be realized
approximately. Principally, then, in this note we aim to
clarify and extend those
earlier works.

Let us consider a naive picture of a black hole of mass $M$ in thermal
equilibrium in $1+1$ dimensions, in a box of length $L$.  Let the radiation in
the box be of massless bosons of temperature $T$.  Here we follow closely the
line of reasoning presented in the initial pages of Ref.~[27].  As suggested
by earlier investigations of the two-dimensional black hole,$^6$ the
temperature of the black hole, $T_{\rm BH}$ is fixed and independent of the
mass.   The total energy and entropy of the system is
$$\eqalign{U &= M + aLT^2\cr
S&= {M\over T_{\rm BH}} + 2aLT \ \ ,\cr}\eqno(1)$$
where $a$ depends on the number of species of bosons in the thermal bath.
Extremizing the entropy at fixed energy yields $T = T_{\rm BH}$ and the second
derivative of the entropy with respect to temperature at this extremum is
negative,
$${d^2S\over dT^2}= {1\over T_{\rm BH}} \  {\partial^2M\over 2T^2}\bigg|_U = -
{2aL\over T_{\rm BH}} < 0 \eqno(2)$$
suggesting that the system is stable.  As we show later, when we write down a
solution to the field equations that correspond to a black hole in a heat bath,
this picture is too naive; in $1+1$ dimensional dilaton gravity the presence
of the bath very dramatically affects the geometry of the spacetime.

To begin with, consider the low-energy effective action of string
propagation in a two-dimensional manifold $({\cal M},g)$
$$\eqalign{I &= I_G + I_M\cr
I_G &= {1\over 2\pi}\int_{\cal M} d^2x\,\sqrt{-g}\, e^{-2\phi} \left[ R +
4(\nabla\phi)^2 + 4\lambda^2\right] + {1\over \pi} \int_{\partial{\cal M}}
e^{-2\phi} K\,d\epsilon \cr
I_M &= {1\over 2}\int_{\cal M}
d^2x\,\sqrt{-g}\, \sum^N_{\Lambda=1} \left( \nabla f_i\right)^2 \cr}\eqno(3)$$
where $K$ is the trace of the second fundamental form on the boundary,  (this
term is necessary for the $I_G$ to depend on the fields $g,\phi$ and their
first
derivatives only) and where $f_i$ are massless scalars minimally coupled to
gravity.

We look for equilibrium ({\it i.e.\/} globally static)
solutions by requiring the metric possess a global time-like Killing vector.
Any two-dimensional metric that
has a global time-like Killing vector may be put in the form,
$$ds^2 = - \Omega^2(r) dt^2 + dr^2 \ \ .\eqno(4)$$
The non-vanishing Christoffel symbols with this metric are
$\Gamma^1_{00} = \Omega\Omega'$ and $\Gamma^0_{10} = \Omega'/\Omega$ where the
prime denotes $\partial/\partial r$.  The scalar curvature is $R =
-2\Omega''/\Omega$ and the most general static and covariantly conserved
tensor has the form
$$T_{\mu\nu} = \left[ \matrix{ A - {\Omega\over\Omega'}A' & {b\over
\Omega}\cr\noalign{\vskip 0.2cm}
{b\over \Omega} & {A\over\Omega^2} \cr}\right]\eqno(5)$$
where $A$ is an arbitrary function of $r$ and $b$ is a constant.  The
equations of motion that follow from Eqs.~(1) are, in general,
$$\eqalignno{
 {1\over \pi} e^{-2\phi} \left(\nabla_\mu\nabla_\nu\phi+g_{\mu\nu} \left(
\left( \nabla\phi\right)^2 - \nabla^2\phi - \lambda^2\right)\right) + {1\over
2}T^{(f)}_{\mu\nu} &= 0 &(6\hbox{a}) \cr
{1\over\pi} e^{-2\phi} \left( R + 4\nabla^2\phi - 4\left(\nabla\phi\right)^2 +
4\lambda^2\right) &= 0 &(6\hbox{b}) \cr
{1\over \sqrt{-g}} \nabla_\mu g^{\mu\nu} \sqrt{-g}\, \nabla_\nu f_i &= 0
&(6\hbox{c}) \cr}$$
where
$$T^{(f)}_{11} = {1\over 2} \sum^N_{i=1} \left[ \left( \partial_r
f_i\right)^2 + {1\over \Omega^2} \left( \partial_t f_i\right)^2 \right]$$
and $T^{(f)}_{\mu\nu}$ is the stress tensor of the fields $f_i$.  Since $D^\mu
T^{(f)}_{\mu\nu}=0$, Eq.~(6a) implies that
$$R+4\nabla^2\phi - 4(\nabla\phi)^2$$
is a constant, so Eq.~(6b), while an independent equation, is consistent with
Eq.~(6a).  Equation (6b) is thus essentially a statement of the gravitational
Bianchi identity.

To look for static solutions to Eq.~(6) with the
metric of Eq.~(4) we learn immediately that $T^{(f)}_{\mu\nu}$ may
not be of the most general form of Eq.~(5).  Indeed, (6a) for $\mu=1$, $\nu=0$
implies that $b=0$.  Since the matter fields $f_i$ that we are minimally
coupling to gravity are massless we expect that (at least classically)
$T^{(f)}_{\mu\nu}$ is traceless.  This forces $A=2A_0$, a constant.
  $A_0$ may be thought of as the local
energy density.  It is easy to check that constant $A$ is consistent
with the
equations of motion for the $f_i$, Eq.~(6c), but we are, for now, not
interested in explicit classical solutions for the $f_i$.
The remaining equations of motion in
this metric {\it ansatz\/} read;
$$\eqalignno{
\phi'' - \left(\phi'\right)^2 + \lambda^2 + {\pi A_0\over \Omega^2} e^{2\phi}
&= 0 &(7\hbox{a}) \cr
\phi'' + {\Omega'\over\Omega}\phi' - \left( \phi'\right)^2 + \lambda^2 -
{\Omega''\over2\Omega} &= 0 &(7\hbox{b}) \cr
- {\Omega'\over\Omega}\phi' + \left( \phi'\right)^2 - \lambda^2 + {\pi A_0\over
\Omega^2} e^{2\phi} &= 0 &(7\hbox{c}) \cr}$$
Eliminating the terms involving $A_0$ permits one to integrate once, finding
$$\phi' = {\Omega'\over\Omega} + {c\over\Omega}\eqno(8)$$
where $c$ is a constant of integration.  These equations of motion are
invariant under together rescaling $\Omega$ and by shifting $\phi$ by a
constant, and so there are only really three possibilities for $c$; either 0
or $\pm\lambda$.  It is simple to show that the solutions with $c = \pm\lambda$
are: a) $\Omega
= \tanh \lambda r$, $\phi =-{\rm ln}(\cosh \lambda r) + \phi_0$; the LDV
$\Omega=1$, $\phi=-\lambda r+\phi_0$; and b)
$\Omega = \cosh \lambda r$, $\phi = {\rm ln}(\sinh
\lambda r)+\phi_0$, respectively.
Regions a) and b)  correspond to those of the maximally extended black hole
solutions$^{10,28-31}$ and a), the LDV, and b) are solutions with $A_0=0$.

For $c=0$, Eqs.~(7) reduce to
$${\Omega''\over 2\Omega} - \left( {\Omega'\over\Omega}\right)^2 + \lambda^2
= 0 \eqno(9)$$
which has the general solution
$${1\over\Omega} = B_+ e^{\sqrt{2}\,\lambda r} + B_- e^{-\sqrt{2}\, \lambda
r}\ \ ,\qquad \phi = {\rm ln}\Omega + \phi_0 \eqno(10)$$
with $B_\pm$, $\phi_0$ being real numbers.

These solutions all have $A_0 = {\lambda^2\over \pi} e^{-2\phi_0}\not=0$.  They
are static spacetimes filled with radiation density, $A_0$.

Before including back-reaction or studying the thermodynamics of these
spacetimes we briefly describe their geometry.  These solutions, Eq.~(10),
all approach
constant curvature $R = -4\lambda^2$ asymptotically, $r\to \pm\infty$.  There
are three different geometries possible from Eq.~(10),
depending on the relative sign of $B_+$ and
$B_-$.  Note that as described above, we may scale $\Omega$ such that, without
loss of generality, $B_+=1$.  Three different geometries result from
whether $B_-$ is positive, zero, or negative. Essentially they are
$$\Omega = \cases{ {1\over 2\cosh \sqrt{2}\, \lambda r} &\qquad $B_->0$\qquad
(I)\cr\noalign{\vskip 0.2cm}
e^{-\sqrt{2}\, \lambda r} &\qquad $B_-=0$\qquad (II) \cr\noalign{\vskip 0.2cm}
{1\over 2\sinh \sqrt{2}\,\lambda r} &\qquad $B_-<0$\qquad (III) \cr}$$

Solutions I and II are geodesically complete and III has a naked singularity
at $r=0$.  Solution~II is another linear dilaton solution (LDV) $\phi =
- \sqrt{2}\,\lambda r+\phi_0$.  Solution~II is a constant curvature metric and
therefore has, in addition to the trivial time translation Killing vector,
two other Killing vectors, one associated with translations in $r$ and the
other akin to a boost. This is in strong
analogy to the LDV solution (flat space) with $A_0=0$.
It is instructive to compare these solutions with the black hole
solution.  In null coordinates $x^+$ and $x^-$
the black hole metric (with $A_0 = 0$) is$^{6,\,10}$
$$ds^2_{\rm BH} = -{dx^+ dx^- \over {\displaystyle{M\over\lambda}} -
\lambda^2x^+ x^-}\ \ .\eqno(11)$$
In the coordinates
$$x^\pm = \sqrt{{M\over 4\lambda}} \left[ {\sqrt{2}\over \lambda} \cosh
\sqrt{2}\, \lambda r\pm t\right]$$
the metric of spacetime III is
$$ds^2_{\rm III} = - {dx^+ dx^-\over {\displaystyle{M\over\lambda}} -
{\displaystyle{\lambda^2\over 2}}\left( x^+ + x^-\right)^2} \ \
.\eqno(12)$$
The classical solution in conformal gauge has the
general solution
$$ds^2 = - e^{2\rho} dx^+ dx^- \eqno(13)$$
with
$$\partial_+ \partial_- e^{-2\rho} = - \lambda^2 \ \ ,\quad
\partial_+\partial_+ e^{-2\rho} = - t_+ (x^+)\ \ ,\quad \partial_- \partial_-
e^{-2\rho} = - t_- (x^-)\eqno(14)$$
where $t_\pm(x^\pm)$ are the right and left moving parts of the
$T^{(f)}_{\mu\nu}$. Thus, solutions I and III may be interpreted as a
two-dimensional dilatonic black hole in a radiations bath {\it where\/} the
naive temperature of the black hole $\left( T_{\rm BH} \sim {\lambda\over
2\pi}\right)$ is the same as that of the radiation bath.  This is quite
different than known solutions to general relativity in four dimensions where
classically there are static cosmological solutions with $T_{\rm BH}>T_{\rm
cosmo}$ (see Refs.~[32,33]), where $T_{\rm cosmo}$ is a cosmological
temperature associated with the cosmological horizon.  In four dimensional
Einstein gravity it
seems technically formidable to find analytic solutions corresponding
to a black hole in
equilibrium with a heath bath. We will remark on the relation
of the solutions of the two-dimensional model presented here and
ordinary gravity in four dimensions later.

The interpretation of spacetimes I and III as the
two-dimensional dilaton black hole in a radiation bath is, although correct,
not quite the naive model of the black hole "in-a-box" discussed
after the introduction.
The presence of uniform, static
energy density $A_0$ drastically modifies the geometry of the spacetime, and
there is no smooth limit as $A_0\to 0$; the equations with and without the
bath ($A_0=0$ and $A_0\not=0$, respectively) are very different.  It is still
useful for taxonomic purposes to regard the solutions I, II and III as
related to a), the LDV and b) respectively, of the black hole spacetime
but including the effects of a radiation bath.\footnote{*}{Note although
regions a) and b) (see discussion after Eq.~(8))
of the black hole metric are related by
duality,$^{29-31,38}$ when matter is included, duality is no longer a
symmetry; there is no duality symmetry relating solutions I and III.}
 For example, the curvature scalar of I is positive near
$r\approx 0$ while that for III is strictly more negative than $-4\lambda^2$.
Also note that spacetimes I and III only have a single global Killing vector.
That is, the $r$-translation isometry of spacetime II is broken spontaneously
in spacetimes I and III, as expected of black hole backgrounds.  Later, in
computing asymptotic masses we will again see how natural it will be to think
of the solutions I, II and III as corresponding to the patches a), the LDV and
b), respectively, of the black hole spacetime.

It is easy to compute the Hawking radiation$^{12}$ of, and back-reaction on
these spacetimes.  Following
Refs.~[8,10] this may be done in two steps.   First, Hawking radiation arises
from the conformal anomaly of the matter fields,$^{34,35}$
$$\ll T_{\mu\nu} g^{\mu\nu}\rr = - {N\over 12} R \eqno(15)$$
Now, following Ref.~[10] and using metrics I, II and III, one may easily
compute semiclassically what the Hawking flux at $r=\infty$.  It is simple
to show that this flux is negligibly small compared to the ambient matter
($A_0\not=0$) at infinity.  Thus spacetimes I,II and III
do not semiclassically decay,
unlike the black hole without a radiation bath ($A_0=0$ and with
trivial boundary conditions at $r=\infty$.)
Furthermore, due to the fact that these metrics are
functions of $\left(x^++x^-\right)$, we find that
at infinity $x^+\to\infty$ or $x^-\to\infty$
the components of the full renormalized stress tensor
$T_{++} \left(x^+,x^-\right)$ and
$T_{--}\left(x^+,x^-\right)$
are  equal
% $\forall x^+,x^-$
and in the limit
proportional to $A_0$.

Of course it may be argued that there is nothing really ``asymptotic'' about
$r\sim\infty$.  It is simple to show that for the spacetimes I, II, III a
time-like trajectory reaches $r=\infty$ in a finite proper time proportional
to $1/\lambda$.  Indeed it is easy to find coordinates in which one may study
the maximal extensions of spacetimes I, II and III.  Figure~1 contains the
Penrose diagrams for these spacetimes.  It is natural to regard spacetime II
as the ``vacuum'' to which we should compare the solutions I and III, as was
the LDV for the black hole without a radiation bath $(A_0=0)$.  In this sense,
``asymptotic'' will mean $r\sim\infty$ since there the solutions I and III
approach solutions II.

It is somewhat more interesting to study the back-reaction of the Hawking
radiation in these solutions. For the black hole it was hoped that studying
the back-reaction would further clarify gravitation collapse, subsequent
evaporations, and the problem of information loss.

In the metric {\it ansatz\/}, Eq.~(4), the most general, static, covariantly
conserved stress tensor (see Eq.~(5)) satisfying Eq.~(15) is:
$$T_{\mu\nu} = \left[ \matrix{ 2A_0 - \alpha\left({\left(
\Omega'\right)^2\over 2} - \Omega\Omega''\right) & {b\over \Omega}
\cr\noalign{\vskip 0.2cm}
{b\over\Omega} & {2A_0\over \Omega^2} - {\alpha\over 2} \left( {\Omega'
\over\Omega}\right)^2 \cr} \right] \eqno(16)$$
where $A_0$ and $b$ are again constants and $\alpha=N/12$.

To understand how the Hawking radiation back-reacts on the metrics I, II and
III, we can begin by putting Eq.~(16) into Eq.~(6a).  This gives the equations
of motion with back-reaction,
$$\eqalignno{
\phi'' - \left( \phi'\right)^2 + \lambda^2 + {\pi e^{2\phi}\over \Omega^2}
\left( A_0 - {\alpha\over 2}\left( {\left(\Omega'\right)^2 \over 2} -
\Omega\Omega''\right)\right) &= 0 &(17\hbox{a}) \cr
\phi'' + {\Omega'\over\Omega}\phi' - \left( \phi'\right)^2 + \lambda^2 -
{\Omega''\over 2\Omega} &= 0 &(17\hbox{b}) \cr
- {\Omega'\over \Omega}\phi' + \left( \phi'\right)^2 - \lambda^2 + {\pi
e^{2\phi}\over\Omega^2} \left( A_0 - \alpha{\left(\Omega'\right)^2\over
4}\right)&=0  &(17\hbox{c}) \cr}$$

Note that for metric II, the LDV solution,
the terms in Eq.~(17) due to the back-reaction
are indeed negligibly small in the $r\to \infty$ region.  This is also
true for the other solutions (I,III).  Thus our notion of ``asymptotic,'' as
described above for these solutions is unchanged by
including the back-reaction.
Although for $A_0=0$ the LDV solution remains a
solution even including back-reaction, for $A_0\not=0$ it is simple to show
that there can be no exactly LDV solution (no solution with $\phi = -
\beta r + \phi_0$ for some $\beta$).

The asymptotic solution to Eq.~(17) are, $(r\to\infty)$
$$\eqalign{ \phi(r) &\to \sqrt{2}\,\lambda r + \phi_0 + e^{-2\sqrt{2}\,\lambda
r} \left( \omega_0 + {2\alpha\over A_0} \left( {\sqrt{2}\over 3}\lambda r -
{1\over 6}\right)\right) \cr
\Omega(r) &\to e^{-\sqrt{2}\,\lambda r} \left( 1 + e^{-2\sqrt{2}\,\lambda r}
\left( \omega_0 + \alpha {2\sqrt{2}\over 3} \lambda r\right)\right)
\cr}\eqno(18)$$
where $\omega_0<0$, $\omega_0=0$, $\omega_0>0$ correspond to asymptotic views
of
the solutions I, II and III, respectively.

Although no exact solution to
Eq.~(17) with $A_0 \not=0$ (or $A_0=0$) is known, we can ascertain many
properties of any solution.  For example, letting $y = e^{-2\phi}+\pi\alpha$,
it is straightforward to show that Eq.~(17) imply
$$\eqalign{
\left[4\lambda^2 + {\Omega'\over\Omega}\left( {\Omega\Omega''\over\left(
\Omega'\right)^2} \right)' ~\right] y &= \pi \left( 4\alpha\lambda^2 -
{\alpha\Omega''\over\Omega} + 4A_0 {\Omega''\over\Omega \left(
\Omega'\right)^2}\right) \cr
\left( \Omega'y\right)' &= - {2\pi\over\Omega} \left( A_0 -
{\alpha\left(\Omega'\right)^2\over 4}\right)\ \ .\cr}\eqno(19)$$
We now study these equations near a power law singularity, such as that found
in spacetime III.
For the metric to behave as
$$\Omega \sim r^\gamma\ \ ,\qquad \hbox{as}\ \ r\to 0\qquad (\gamma\not=0,\ \
\hbox{real})\eqno(20)$$
(which could indicate either a horizon $\gamma>0$ or a singularity $\gamma<0$)
using Eq.~(19) we find that including back-reaction only allows $\gamma=1$.
That is, roughly speaking, including back-reaction is not consisent with
simple power-law
singularities in the metric.  Of course, Eq.~(15) is a one-loop result and
near curvature singularities we might expect that there may be other
contributions to $\ll T_{\mu\nu} g^{\mu\nu}\rr$ that would dominate.  By
``singularity'' we really mean curvatures approaching ``1'' in units of
$\lambda^2/\alpha$.  As a consequence of this we see that including back
reaction will strongly modify the singularity at $r=0$. This
is also true for the singularity of the black hole in the absence of a bath
($A_0=0$).

Similarly one may show that back-reaction also strongly modifies solutions
II in the strong coupling region $r\to-\infty$. However, for
metric I the corrections due to including the back-reaction appear to be mild
almost everywhere.  It can also
be shown that Eqs.~(19) are consistent with $\Omega'$
vanishing linearly.  Since this is the distinctive feature of the ``throat''
region ($r\approx 0$) of the metric in I, this indicates that there may be an
exact semiclassical solution (${i.e.}$ with back-reaction included)
that is qualitatively
metric I.

We now associate ``asymptotic'' masses to the spacetimes discussed above. This
will support the notion that spacetimes I, II, and III should be thought of as
a), the LDV and b), respectively but with spacetime filled with a radiation
bath.

Before continuing we note that, as described earlier, we have made a special
choice in defining what we mean by ``asymptotic.''  With this definition we
are not {\it a priori\/} guaranteed that all ``definitions'' of mass will lead
to the same expressions.  Below we discuss a thermodynamic
definition of ``asymptotic' mass. However,
first, we would like to point out that
one may compute an ``asymptotic'' mass by simply comparing the geodesics
near
$r=\infty$ of, say, spacetime I to those of the ``vacuum'' spacetime II.
Recall the ``vacuum'' spacetime II is anti-deSitter space and, as usual,
we consider geodesics in the covering space shown in Fig.~1.  One may compute
the ``transit time'' to cross a wedge: that is,
the proper time for an observer
to enter and leave the region marked II.  As expected, it is the same for all
time-like trajectories and is $\sim 1/\lambda$.  Now, when one solves the
geodesic equation for geodesics near $r\sim\infty$ in spacetime I, one finds
that their transit time is a little bit slower than that of spacetime II.
This difference is attributable to an extra overall attractive mass
$M\sim A_0/\lambda$ in spacetime I as compared with spacetime II.
The natural position to associate to this mass
would be near $r=0$ in spacetime I, since that
is where the metric deviates appreciable from the "vacuum" metric
of spacetime II.
Again, one may even more simply compare the instantaneous
accelerations of a body near $r\to\infty$ but initially at rest in both
spacetimes I and II.  One again finds that the extra acceleration towards $r=0$
in space I may be attributable to a mass $M\sim A_0/\lambda$. These
masses are to be thought of as contravariant quantites with respect to
the metric II. It should also be possible to derive an
ADM-like$^{6,37}$ mass formula
directly from the equations of motion, but we do not pursue that here.

We now compute the equilibrium thermodynamic functions for the spacetimes I,
II and III.  This will give a deeper understanding of
these solutions and will be convenient for discussing issues of stability and
fluctuation below.  In what follows we proceed essentially from
Refs.~[28,36].
%and show in the Appendix
%that the ADM$^{6,37}$ formalism yields the same asymptotic
%mass.\footnote{*}{In the ADM formalism, it is also quite straightforward to
%include the effects of back reactions.  This is described in the Appendix.}

To understand thermodynamic properties of the spacetime semiclassically, it
is most convenient to compute the free energy.  This is done by simply
evaluating the action Eq.~(3) on a Euclidean continuation of the solutions in
question. In our metric {\it ansatz\/} Eq.~(4), this yields
$$I = {1\over \pi} \int_{\partial{\cal M}}
e^{-2\phi}\left( {\Omega'\over\Omega} -
2\phi'\right) d\Sigma + B_{\rm mat} \eqno(21)$$
where we have computed the action in some box, the boundary $\partial{\cal M}$
of which is located at some fixed parameter distance $r_w$.  $B_{\rm mat}$
is a bulk term due entirely to the $I_M$ of the radiation bath.

Call $T_w$ the temperature that is seen by an inertial observer at the
wall.  Tolman's relation (which is a consequence of thermal equilibrium)
indicates,
$$T_w = {\tilde T\over \Omega(r_w)} \eqno(22)$$
where $\tilde T$ is some constant temperature. Later it will be identified
with the temperature of the ambient radiation. This relation implies that in
the Euclidean continuation of these spacetimes, the $t$-direction of the
manifold at $r=r_w$ is periodic, with extent $\Omega/\tilde T$.  Thus we
find that, for fixed $A_0$, $B_{\rm mat}$ is the same linear function in
$r_w$ for any solution.  We are fundamentally interested in
distinguishing the solutions I, II and III from one another and so one can
show $B_{\rm mat}$, the bulk term, may be ignored; differences between the
solutions will be manifest in differences in their $I_G$ only.

Following Ref.~[28] it is useful to consider the free energy $F$ as a
function of $T_w$ and $D$, the total dilaton charge within the box.  $D$
may be conveniently defined as the charge associated with the current
$$j^\mu = \epsilon^{\mu\nu} \partial_\nu\, e^{-2\phi}\eqno(23)$$
where $\epsilon^{\mu\nu}$ is the antisymmetric covariant tensor
($\epsilon_{01} = \sqrt{-g}$) so $\nabla_\mu j^\mu =0$.  The charge within
the box is thus
$$D = \int ^{r_w} j_a d\Sigma^a = e^{-2\phi_w}\eqno(24)$$
where $\phi_w=\phi(r_w)$ is the value of the dilation field at the wall.

It is now easy to compute the free energy $F = - T_w I$ from Eq.~(21).
For the spacetime II,
$$F^{\rm II} = - {\sqrt{2}\, \lambda\over \pi} D \eqno(25)$$
and so, the entropy in the gravitational field is
$$S^{\rm II} = - {\partial F^{\rm II}\over \partial T_w}\bigg|_D = 0 \ \
.$$
As in the case without a radiation bath $(A_0=0$), the pure LDV solution has
zero entropy but finite energy $U^{\rm II} = F^{\rm II} + T_w S^{\rm II}
= F^{\rm II}\not=0$.  This may seem curious for a vacuum solution, but is
expected since this is energy tied up in the dilaton field.

We now compute the free energy of spacetime I.
With
$$D = 4e^{-2\phi_0} \cosh^2\left(\sqrt{2}\, \lambda r_w\right)\ \ ,
\qquad T_w = T
\cosh \left(\sqrt{2}\,\lambda r_w\right)\eqno(26)$$
we find the free energy of spacetime I to be
$$F^{\rm I} = - {\sqrt{2}\,\lambda DT\over \pi T_w}\sqrt{\left( {T_w\over
T}\right)^2-1} \eqno(27)$$
where $T$ is a constant that we will relate to the temperature of the
radiation bath (see Eq.~(22)).  Again using $S = - {\partial F\over \partial
T_w}\bigg|_D$ we find the entropy to be
$$S^{\rm I} = {\sqrt{2}\,\lambda DT\over \pi T^2_w \sqrt{ \left(
{\displaystyle{T_w\over T}}\right)^2-1}}\ \ .\eqno(28)$$
Note that the entropy is positive and that its ``asymptotic'' value is zero,
which is consistent with the fact that the metric of I has no horizon.  The
energy $U^{\rm I} = F^{\rm I} + T_w S^{\rm I}$ is
$$U^{\rm I} = - {\sqrt{2}\,\lambda D\left[ \left( {\displaystyle{T_w\over
T}}\right)^2 - 2 \right] \over \pi
\left(\displaystyle{T_w\over T}\right) \sqrt{ \left(
{\displaystyle{T_w\over T}}\right)^2-1} }\ \ .\eqno(29)$$
Thus, an ``asymptotic'' observer would ascribe a mass to this spacetime,
$$M^{\rm I} = \lim_{r_w\to\infty} \left( U^{\rm I} - U^{\rm II}\right) =
{6\sqrt{2}\, \lambda e^{-2\phi_0}\over \pi} = {6\sqrt{2}\,A_0\over\lambda}
\eqno(30)$$
where we have made use of the fact that $e^{-2\phi_0} = {\pi
A_0\over\lambda^2}$.  Thus, as in the black hole case ($A_0=0$) we again find
that the mass of the spacetime is related to the constant term of the dilaton.
This reinforces the interpretation of metric I as that of the two-dimensional
dilaton black hole in equilibrium with a radiation bath.  It also concurs with
the comparison of the transit times described in the beginning of this
section.

The fact that this analysis has yielded an asymptotic mass for a spacetime
that asymptotically has zero entropy seems to be at odds with the expectations
that $S = M/T$.$^{28,\,26}$  This is simply due to the fact that the
quantities $F$, $M$, $T_w$ are really to be thought of as contravariant
({\it i.e.\/} possessing an upper index) quantities while $S$ and $T$ are
really globally defined (scalar quantities).  Roughtly speaking, if the
``asymptotic'' region is
 not simply flat space, general covariance suggests one should
expect a relation rather like $S = M/T_w$.  Thus $S$ may vanish although
$M$ is non-zero.

A final aside:  Were we to repeat this computation of the mass for spacetime
III we would find $M_{\rm III} = - 2\sqrt{2}\,A_0/2\lambda$.  Negative mass
solutions are not necessarily
excluded since the dilaton contributes to the equations of
motion in such a way that the dominant energy condition is violated and so the
mass of spacetime is not bounded below.
For spacetime III, it is related to the fact that this spacetime has a
naked singularity.
This identification of the mass is
also  consistent with the taxonomy described earlier, that we really should
think of I, II and III as region a), LDV, b) of the extended black hole
solution but in  a bath of radiation.

It is straightforward to go beyond the static {\it ansatz\/} Eq.~(4) and
study the full linear response of the equations of motion.  Linearizing
Eq.~(6) reveals that the static solutions are stable only to time-dependent
perturbations that vanish spatially at least as $\Omega^4$ in the asymptotic
region.  However, such perturbations seem not to contribute to the
asymptotic mass, and so do not correspond to throwing mass into the ``hole''
(neck at $r=0$). Time-dependent perturbations larger than this represent
spacetimes that are asymptotically not constant curvature.  Indeed there are
two possible scenarios with regard to ``throwing mass'' into the ``hole'' in
these spacetimes.  Either the perturbation causes a shift in the dilaton,
thereby increasing the mass of the spacetime {\it or\/} it causes collapse
to ensue in which some of the radiation energy of the bath falls
,with the perturbation,
into the hole, perhaps
forming a singularity.  A cursory analysis indicates that the
latter possibility occurs, but more work remains to be one.  For a related
discussion in four dimensions, see Refs.~[27,39].

More fundamentally, the study of equilibrium spacetimes including effects like
Hawking radiation should include a discussion of the equilibrium power
spectrum of noise in the spacetime.  The most natural way to do this is by
treating the thermal bath quantum mechanically and understanding Hawking
radiation as dissipation of (gravitational) energy from the geometry of the
space to the bath.  For a stable, equilibrium spacetime it should be
possible to understand the coupling between the hole and the bath that is
represented by the Hawking radiation in terms of
linear transport coefficients.   At
this time it seems that to compute linear transport coefficients one would
need a clearer picture (for example, a realization of Hawking radiation as
arising from a Hamiltonian in the bath's phase-space co-ordinates) of
Hawking's effect than the cursory one presented here.  Nonetheless,
the power spectrum of noise could be expressed in terms of these transport
coefficients.  Work is underway to ascertain how
complete this
view of equilibrium is
for the spacetimes presented here and their cousins in
four dimensions.
\goodbreak
\bigskip
\noindent{\bf CONCLUSION AND SPECULATIONS}
\medskip
We have found static solutions of two-dimensional dilaton gravity coupled to
matter that may be interpreted as a two-dimensional black hole at equilibrium
in a bath of radiation.  We have computed the thermodynamic potentials and
identified the mass attributable to the black hole.  We have also included
the Hawking radiation's back-reaction on the metric and dilaton in these
solutions.

Of course, this toy model of gravitations is somewhat different than
higher-dimensional Einstein--Hilbert gravity. However, we feel some ideas
discussed here may be fruitfully explored in Einstein--Hilbert
gravity in four dimensions.

We conclude with one intriguing speculation. There are many issues
surrounding the final stages of black hole evaporation.  Information loss and
related questions
aside, it is valid to ask simply how would one describe the final
stage of the evaporation. Note that our spacetime I has two interesting
properties: it has no horizon and the ``mean density'' ($\sim M\lambda$) of the
``hole'' is equal to the energy density of the bath $A_0$.  This is
essentially due to the fact that in two dimensions Newton's constant is
dimensionless.  Our intuition would suggest that, for four-dimensional black
holes, immersing the black hole in a radiation bath at the same temperature as
that of the hole would make neither the horizon evaporate nor tame the
singularity.  This results from the fact
that for macroscopic black holes in four dimensions, the radiant energy
density at the hole's temperature is much smaller than the ``mean density'' of
the hole.  As evaporation proceeds, and
somewhat before one gets to temperatures of the
Planck scale, the
equilibrium radiation energy density outside the hole becomes
comparable to
the ``mean density'' of the hole itself.
Furthermore, near the endpoint of the evaporation,
since the area of the horizon is so minute
and the radiation density is so high, any single (interacting) quanta would
likely see many radiation lengths of other quanta
between it and
%to get through on its way to
spatial infinity.  Thus locally, near the evaporating hole, something akin to
a quasi-equilibrium state (with respect to time scales of ${\cal O}\left(
1/M_p\right)$) may be achieved.  In such a state in analogy to the solution I
presented here, we conjecture the horizon may ``evaporate'' and the
singularity melt away.  It would be interesting to investigate the validity of
this ``picture'' of the final stage of black hole evaporation in a more
realistic model.

\goodbreak
\bigskip
\centerline{\bf ACKNOWLEDGEMENTS}
\medskip
The author wishes to thank D.~Freed, D.~Z.~Freedman, M. ~Ortiz
and M.~D.~Perry for
conversations and assistance and D.~Cangemi, M.~Leblanc, N.~Rius and
I.~M.~Singer for encouragement and support.
\goodbreak
\bigskip
\centerline{\bf NOTE ADDED}
\medskip
After completion of this project, the author received two papers,
Refs.~[40,41] which describe cosmological solutions that are somewhat related
to the static spacetimes discussed here.
\vfill
\eject
\centerline{\bf REFERENCES}
\medskip
\item{1.}K. Bardakci and M. B. Halperin, {\it Phys. Rev.\/} {\bf D3}, 2493
(1971).
\medskip
\item{2.}P. Goddard, A. Kent and D. Olive, {\it Phys. Lett.\/} {\bf B152}, 88
(1985).
\medskip
\item{3.}K. Bardakci, M. Crescimanno and E. Rabinovici, {\it Nucl. Phys.\/}
{\bf B344}, 344 (1990).
\medskip
\item{4.}G. Mandal, A. Sengupta and S. R. Wadia, {\it Mod. Phys. Lett.\/} {\bf
A6}, 1685 (1991).
\medskip
\item{5.}S. Elitzur, A. Forge ad E. Rabinovici, {\it Nucl. Phys.\/} {\bf
B359}, 581 (1991).
\medskip
\item{6.}E. Witten, {\it Phys. Rev.\/} {\bf D44}, 314 (1991).
\medskip
\item{7.}G. T. Horowitz, USCBTH--92--32, HEPTH/9210119.
\medskip
\item{8.}J. A. Harvey and A. Strominger, EFI-92-41, HEPTH/9209055.
\medskip
\item{9.}S. B. Giddings and A. Strominger, UCSB-TH-92-01, HEPTH/9202004.
\medskip
\item{10.}C. Callan, S. B. Giddings, J. A. Harvey and A. Strominger, {\it
Phys. Rev.\/} {\bf D45}, R1005 (1992).
\medskip
\item{11.}B. Birner, S. B. Giddings, J. A. Harvey and  S. Strominger,
UCSB-TH-92-08; EFI-92-16; HEPTH/9203042.
\medskip
\item{12.}S. W. Hawking, {\it Phys. Rev.\/} {\bf D14}, 2460 (1976).
\medskip
\item{13.}S. B. Giddings and W. M. Nelson, UCSBTH-92-15, HEPTH/9204072.
\medskip
\item{14.}A. Peet, L. Susskind and L. Thorlacius, {\it Phys. Rev.\/} {\bf
D46}, 3435 (1992).
\medskip
\item{15.}J. G. Russo, L. Susskind and L. Thorlacius, {\it Phys. Lett.\/} {\bf
B292}, 13 (1992).
\medskip
\item{16.}L. Susskind and L. Thorlacius, {\it Nucl. Phys.\/} {\bf B382}, 123
(1992).
\medskip
\item{17.}A. Bilal and C. Callan, PUPT-1320, HEPTH/9205089.
\medskip
\item{18.}S. P. DeAlwis and J. Lykken, Fermilab-Pub-91/198-7.
\medskip
\item{19.}S. P. DeAlwiss, COLO-HEP-284, HEPTH/9206020;
COLO-HEP-288, HEPTH/9207095; COLO-HEP-280, HEPTH/9205069.
\medskip
\item{20.}A. Mikovic, QMW/PH/92/16, HEPTH/9211082.
\medskip
\item{21.}U. H. Danielsson, CERN-TH.6711/92, HEPTH/9211013.
\medskip
\item{22.}K. Hamada, UT-Komaba 92-9, HEPTH/9210101.
\medskip
\item{23.}S. B. Giddings, UCSBTH-92-36, HEPTH/9209113.
\medskip
\item{24.}S. B. Giddings and A. Strominger, UCSBTH-92-28, HEPTH/9207034.
\medskip
\item{25.}A Strominger, UCSBTH-92-18, HEPTH/9205028.
\medskip
\item{26.}V. P. Frolov, in {\it Quantum Mechanics in Curved Space-Time\/}, J.
Audretsch and V. de~Sabbata, eds. (Plenum Press, NY, 1990), p.~141.
\medskip
\item{27.}G. W. Gibbons and M. J. Perry, {\it Proc. R. Soc. London\/} {\bf
A358}, 467 (1978).
\medskip
\item{28.}G. W. Gibbons and M. J. Perry, HEPTH/9204090.
\medskip
\item{29.}A. Giveon, {\it Mod. Phys. Lett.\/} {\bf A6}, 2843 (1991).
\medskip
\item{30.}E. Kiritsis, {\it Mod. Phys. Lett.\/} {\bf A6}, 2871 (1991).
\medskip
\item{31.}A. A. Tseytlin, {\it Mod. Phys. Lett.\/} {\bf A6}, 1721 (1991).
\medskip
\item{32.}L. F. Abbott and S. Deser, {\it Nucl. Phys.\/} {\bf B195}, 76
(1982).
\medskip
\item{33.}G. W. Gibbons and S. W. Hawking, {\it Phys. Rev.\/} {\bf D15}, 2738
(1977).
\medskip
\item{34.}P. C.W. Davies, S. A. Fulling and W. G. Unruh, {\it Phys. Rev.\/}
{\bf D13}, 2720 (1976).
\medskip
\item{35.}S. M. Christensen and S. A. Fulling, {\it Phys. Rev.\/} {\bf D15},
2088 (1977).
\medskip
\item{36.}G. W. Gibbons and S. W. Hawking, {\it Phys. Rev.\/} {\bf D15}, 2752
(1977).
\medskip
\item{37.}R. Arnowitt, S. Deser and C. W. Misner, ``The Dynamics of General
Relativity,'' in {\it Gravitation: An Introduction to Current Research\/}, L.
Witten, ed. (J. Wiley \& Sons, Inc., New York, 1962).
\medskip
\item{38.}E. Kiritsis, LPENS-92-30, HEPTH/9211081.
\medskip
\item{39.}D. J. Gross, M. J. Perry and L. G. Yaffe, {\it Phys. Rev.\/} {\bf
D25}, 330 (1982).
\medskip
\item{40.}F. D. Mazzietelli and J. G. Russo, UTTG-28-92, HEPTH/9211095.
\medskip
\item{41.}M. Yoshimura, TU/92/416.
\vfill
\eject
\vskip .3in
\noindent{\bf Figure 1} ~~The physical maximally
extended spacetime diagrams of I, II and III ~~
(There is no physical reason to extend spacetime beyond a
strong coupling region.)
\vfill
\end